\documentclass[sigconf]{acmart}
\AtBeginDocument{%
  }

\setcopyright{acmlicensed}
\acmYear{2025}\copyrightyear{2025}
\acmConference[CPSS '25]{11th ACM Cyber-Physical System Security Workshop}{August 25--29, 2025}{Hanoi, Vietnam}
\acmBooktitle{11th ACM Cyber-Physical System Security Workshop (CPSS '25), August 25--29, 2025, Hanoi, Vietnam}
\acmDOI{10.1145/3709017.3737711}
\acmISBN{979-8-4007-1413-9/2025/08}

\usepackage[acronym, shortcuts]{glossaries-extra}
\glssetcategoryattribute{acronym}{nohyper}{true}
\setabbreviationstyle[acronym]{long-short}
\newacronym{vm}{VM}{Virtual Machine}
\newacronym{ics}{ICS}{Industrial Control System}
\newacronym{ot}{OT}{Operational Technology}
\newacronym{it}{IT}{Information Technology}
\newacronym{cps}{CPS}{Cyber-Physical System}
\newacronym{plc}{PLC}{Programmable Logic Controller}
\newacronym{hmi}{HMI}{Human Machine Interface}
\newacronym{dos}{DoS}{Denial of Service}
\newacronym{rat}{RAT}{Remote Access Trojan}
\newacronym{opc}{OPC}{OLE for Process Control}
\newacronym{sis}{SIS}{Safety Instrumented System}
\newacronym{snr}{SNR}{Signal-to-Noise-Ratio}
\newacronym{dt}{DT}{Decision Tree}
\newacronym{lr}{LR}{Logistic Regression}
\newacronym{ab}{AB}{AdaBoost}
\newacronym{lof}{LOF}{Local Outlier Factor}

\newacronym{ai}{AI}{Artificial Intelligence}
\newacronym{asr}{ASR}{Attack Success Rate}
\newacronym{ads}{ADS}{Anomaly Detection System}
\newacronym{can}{CAN}{Controller Area Network}
\newacronym{cnn}{CNN}{Convolutional Neural Network}
\newacronym{cpu}{CPU}{Central Processing Unit}
\newacronym{crc}{CRC}{Cyclic Redundancy Check}
\newacronym{ddos}{DDoS}{Distributed Denial of Service}
\newacronym{dnn}{DNN}{Deep Neural Network}
\newacronym{far}{FAR}{False Acceptance Rate}
\newacronym{fas}{FAS}{Fingerprint-based Authentication Systems}
\newacronym{fgsm}{FGSM}{Fast Gradient Sign Method}
\newacronym{fn}{FN}{False Negative}
\newacronym{fp}{FP}{False Positive}
\newacronym{gan}{GAN}{Generative Adversarial Network}
\newacronym{gps}{GPS}{Global Positioning System}
\newacronym{gpu}{GPU}{Graphics Processing Unit}
\newacronym{gru}{GRU}{Gated Recurrent Unit}
\newacronym{ids}{IDS}{Intrusion Detection System}
\newacronym{imu}{IMU}{Inertial Measurement Unit}
\newacronym{ml}{ML}{Machine Learning}
\newacronym{dl}{DL}{Deep Learning}
\newacronym{obd}{OBD}{On-Boad Diagnostic}
\newacronym{oem}{OEM}{Original Equipment Manufacturer}
\newacronym{rnn}{RNN}{Recurrent Neural Network}
\newacronym{shap}{SHAP}{SHapley Additive exPlanations}
\newacronym{svdd}{SVDD}{Support Vector Domain Description}
\newacronym{svm}{SVM}{Support Vector Machine}
\newacronym{tn}{TN}{True Negative}
\newacronym{tp}{TP}{True Positive}
\newacronym{tpm}{TPM}{Trusted Platform Module}
\newacronym{ubm}{UBM}{Universal Background Model}
\newacronym{ubi}{UBI}{Usage Based Insurance}
\newacronym{uds}{UDS}{Unified Diagnostic Services}
\newacronym{mitm}{MitM}{Man-in-the-Middle}
\newacronym{rpm}{RPM}{Revolutions Per Minute}
\newacronym{rf}{RF}{Random Forest}
\newacronym{gmm}{GMM}{Gaussian Mixture Model}
\newacronym{knn}{kNN}{k-Nearest Neighbors}
\newacronym{xai}{XAI}{Explainable Artificial Intelligence}
\newacronym{rl}{RL}{Reinforcement Learning}
\newacronym{nn}{NN}{Neural Network}
\newacronym{if}{IF}{Isolation Forest}
\newacronym{oc-svm}{OC-SVM}{One-Class Support Vector Machine}
\newacronym{apt}{APT}{Advanced Persistent Threat}
\newacronym{vrae}{VRAE}{Variational Recurrent Auto Encoder}

\usepackage{pifont}
\usepackage{subcaption}
\usepackage{graphicx}
\usepackage{fontawesome5}
\newcommand{\yes}{\faCheck}
\newcommand{\no}{\faTimes}
\usepackage{tablefootnote}

\newcommand{\newframework}[0]{\emph{SimProcess}}

\begin{document}

\title[SimProcess: High Fidelity Simulation of  Noisy ICS Physical Processes]{
SimProcess: High Fidelity Simulation of  Noisy \\ ICS Physical Processes
}

\author{Denis Donadel}
\email{denis.donadel@univr.it}
\orcid{0000-0002-7050-9369}
\affiliation{%
  \institution{University of Verona}
  \city{Verona}
  \country{Italy}
}

\author{Gabriele Crestanello}
\email{gabriele.crestanello@studenti.unipd.it}
\orcid{0009-0006-1518-4465}
\affiliation{%
  \institution{University of Padua}
  \city{Padova}
  \country{Italy}
}

\author{Giulio Morandini}
\email{giulio.morandini@studenti.unipd.it}
\orcid{0009-0006-3200-5334}
\affiliation{%
  \institution{University of Padua}
  \city{Padova}
  \country{Italy}
}

\author{Daniele Antonioli}
\email{daniele.antonioli@eurecom.fr}
\orcid{0000-0002-9342-3920}
\affiliation{%
  \institution{EURECOM}
  \city{Sophia Antipolis}
  \country{France}
}

\author{Mauro Conti}
\email{mauro.conti@unipd.it}
\orcid{0000-0002-3612-1934}
\affiliation{%
  \institution{University of Padua}
  \city{Padova}
  \country{Italy}
}

\author{Massimo Merro}
\email{massimo.merro@univr.it}
\orcid{0000-0002-1712-7492}
\affiliation{%
  \institution{University of Verona}
  \city{Verona}
  \country{Italy}
}

\renewcommand{\shortauthors}{Donadel et al.}

\begin{abstract}

Industrial Control Systems (ICS) manage critical infrastructures like power grids and water treatment plants. Cyberattacks on ICSs can disrupt operations, causing severe economic, environmental, and safety issues. For example, undetected pollution in a water plant can put the lives of thousands at stake.  
ICS researchers have increasingly turned to honeypots---decoy systems designed to attract attackers, study their behaviors, and eventually improve defensive mechanisms.
However, existing ICS honeypots struggle to replicate the ICS physical process, making them susceptible to detection.  
Accurately simulating the noise in ICS physical processes is challenging because different factors produce it, including sensor imperfections and external interferences. 
\looseness = -1

In this paper, we propose \newframework{}, a novel framework to rank the fidelity of ICS simulations by evaluating how closely they resemble real-world and noisy physical processes. 
It measures the simulation distance from a target system by estimating the noise distribution with machine learning models like Random Forest. %
Unlike existing solutions that require detailed mathematical models or are limited to simple systems, \newframework{} operates with only a timeseries of measurements from the real system, making it applicable to a broader range of complex dynamic systems. 
We demonstrate the framework’s effectiveness through a case study using real-world power grid data from the EPIC testbed. We compare the performance of various simulation methods, including static and generative noise techniques. Our model correctly classifies real samples with a recall of up to 1.0. It also identifies Gaussian and Gaussian Mixture as the best distribution to simulate our power systems, together with a generative solution provided by an autoencoder, thereby helping developers to improve honeypot fidelity.  
Additionally, we make our code, dataset, and experimental results publicly available to foster research and collaboration. 
\looseness = -1

\end{abstract}



\maketitle

\section{Introduction}

\acp{ics} are crucial for managing and automating critical infrastructure, including power grids, water treatment facilities, and manufacturing plants. These systems integrate hardware and software components to monitor and control physical processes, ensuring efficiency and safety. \ac{ics} interact directly with the physical world, making their security essential to prevent operational disruptions that could lead to economic losses, environmental damage, or safety hazards~\cite{krotofil2013industrial}.

The importance of \ac{ics} security has been highlighted by real-world attacks. A notable example is Stuxnet~\cite{falliere2011w32}, a sophisticated malware that targeted nuclear centrifuges by subtly altering their behavior while remaining undetected by monitoring systems. Other attacks followed, demonstrating how adversaries could exploit \acp{ics} vulnerabilities to manipulate physical processes, causing physical damage~\cite{slowik2019evolution}. This made the research community develop security solutions, such as \ac{ml} based anomaly-detection systems~\cite{mubarak2021anomaly, jamal2023review}, to identify a wide range of attacks. 
\looseness=-1

Recently, \ac{ics} researchers and industries have started to deploy decoy systems to attract attackers and learn about their behavior, the so-called \ac{ics} honeypots. 
Low-interaction honeypots are designed to simulate only a limited set of services or behaviors, typically fooling attackers with minimal interaction to gather data on their tactics~\cite{provos2003honeyd,jicha2016scada}. 
High-interaction  \ac{ics} honeypots emulate a full range of services and behaviors, allowing attackers to engage deeply with the system. 

High interaction honepots can provide comprehensive insights about attacking methods and motivations, but at the cost of greater resource investment and potential risk~\cite{zamiri2019gas, lopez2020honeyplc, lucchese2023honeyics}.
Building a high-interaction honeypot is challenging because it must emulate, among others, industrial control devices, industrial protocols, sensors, actuators, and the underlying physical process. For example, an attacker could spot a honeypot by looking at its software details~\cite{sun2020identifying}.

\paragraph{Motivation.} 
A honeypot developer needs to understand the fidelity that their honeypot can achieve. Fidelity can be increased in every component of the decoy system. For instance, industrial device functionalities can be emulated through high-level computer software, or their firmware can be fully emulated to expose the correct architectures and all the bugs specific to it~\cite{kovavc2023development}. Another important step toward a realistic honeypot is related to the physical process that needs to replicate a real scenario. 

A \emph{realistic physical process simulation} is essential to prevent attackers from understanding that they are dealing with an \ac{ics} honeypot.
In fact, attackers, aware of the presence of honeypots, have started to develop solutions to identify simulated environments. Various techniques have been presented over the years~\cite{srinivasa2023gotta}, but only a few investigated \ac{ics} specific techniques looking at changes in the device registers~\cite{zhu2024honeyjudge} or by employing fuzzy testing~\cite{sun2020identifying}. To our knowledge, the fidelity of industrial physical process simulations has never been investigated thoughtfully.

A central issue with \ac{ics} physical process simulation is reproducing a \emph{noise} consistent with the actual process's characteristics. Such noise may be intrinsic to the physical process because it is generated by its components, including industrial sensors~\cite{makovoz2006noise, dehra2018characterization}, or it may be due to external factors, like weather conditions, electromagnetic interference, or people working on the industrial plant~\cite{ahmed2017hardware}.

However, to our knowledge, no framework exists to assess if a simulated physical process includes realistic noise components. Such a framework would be essential for defenders to build effective \ac{ics} honeypots, even without disclosing the details of the protected physical process. In this scenario, the attacker would hardly detect being in a honeypot and infer information about the real physical process.

\paragraph{Contribution.}

We present \textbf{\newframework}, a framework for benchmarking \ac{ics} physical process simulations and identifying the one that resembles with the highest fidelity a real (noisy) physical process. 
Unlike prior works, our framework only employs a collection of measurements from the real system and does not require the mathematical model (i.e., differential equations) that govern the underlying physical process, allowing us to deal with complex systems, including nonlinear time-invariant (NLTI) ones, and to reproduce variations of the real system. 
\newframework{} extracts selected features from the original signal that are related to the noise, in addition to features extracted from a noise estimation. 

We implemented and evaluated \newframework{} framework on EPIC~\cite{adepu2019epic}, a power grid testbed. Our empirical results demonstrate its capabilities in identifying the best simulation. In particular, our classification models exhibit a recall of up to 1.0 in determining the actual process from various simulations. \newframework{} identified Gaussian, \ac{gmm}, and the autoencoder solutions as the best ways to add noise to resemble the real system and increase the fidelity of the power grid simulation.

We summarize our contributions as follows:
\begin{itemize}
    \item We introduce \newframework{}, the first framework designed to identify simulations that most accurately replicate an ICS physical process by focusing on the characterizing noise.
    \looseness = -1
    \item We implement and evaluate \newframework{} on real-world power grid data coming from EPIC, a state-of-the-art power grid testbed,  and derive simulations of the physical process employing both static additions of noises and generative solutions. 
    \item We show that our models can correctly classify the real physical process with a recall up to 1.0 in a \emph{binary classification} scenario. Moreover, we evaluate the capabilities of \newframework{} of classifying different noises and its ability to preserve the ranking even when dynamic changes are added to the underlying process, making \newframework{} applicable in high-interaction simulations.
    \item We made our code, dataset, and experimental results publicly available: \url{https://github.com/donadelden/SimProcess}
\end{itemize}

\paragraph{Organization.}
Section~\ref{sec:background} briefly introduces important background knowledge, while the system and threat model are presented in Section~\ref{sec:threat}. Section~\ref{sec:metodology} introduces the \newframework{} framework, while Section~\ref{sec:case} presents our case study, the fine-tuning experiments, and final results. Related works are discussed in Section~\ref{sec:related} and Section~\ref{sec:conclusions} concludes this work.

\section{Background}\label{sec:background}

\paragraph{Industrial Control Systems}

An Industrial Control System (ICS) is a system in which industrial operations are supervised, coordinated, controlled, and unified by a computing and communication core~\cite{CPS-DEF}. 
These systems constitute a vital element of \emph{critical infrastructures} that deliver essential services, including water supply, electricity generation and distribution, and nuclear power generation. They combine
industrial hardware and software. 

Industrial processes can be modeled linearly or nonlinearly. With a linear time-invariant (LTI) process, there is a linear relation between an input and the output that does not depend on time. More complex models involve nonlinear terms, time, and (partial) differential equations.

\acp{ics} are highly interconnected. An \ac{ics} network is composed of an enterprise zone containing general purpose \ac{it} systems, connected through a demilitarized zone to a \ac{ot} control zone. The \ac{ot} network operates with the physical process through intelligent devices such as \acp{plc} connected to I/O devices such as sensors and actuators. Operators can supervise and act on the physical process through control systems such as \acp{hmi}~\cite{conti2021survey}. 

\ac{ics} are subject to impactful and large-scale attacks.
A notable attack target example is the power grid, where \ac{ics} controls electricity transmission. The high number of cyberattacks in Ukraine since 2015~\cite{salazar2024tale, shrivastava2016blackenergy}, highlighted the vulnerability of these systems and introduced the need for specific security strategies to protect power systems from cyberattacks~\cite{PowerGrid}. %

\paragraph{ICS Testbeds and Simulations}
\ac{ics} testbeds are experimental environments designed to test and validate technologies, security systems, and protocols in realistic but controlled conditions. They combine physical simulations with real network traffic to assess the security and reliability of critical infrastructures. They could be physical, hybrid, or virtualized~\cite{conti2021survey}. %
While physical testbeds provide high fidelity to the researchers, they are more complex and expensive to develop and maintain. Conversely, a virtualized testbed can also be employed as a decoy system to study attacker's behaviors~\cite{lopez2020honeyplc, lucchese2023honeyics, salazar2024icsnet}.

There are several tools to create virtual \ac{ics} testbed and simulate \ac{ics} physical processes.
MiniCPS~\cite{antonioli2015minicps} or Factory I/O\footnote{Factory I/O - Next-Gen PLC Training (\url{https://factoryio.com/})} are designed for generic \ac{cps} simulation and emulation. Others investigate specific sectors. NEFICS~\cite{salazar2024tale}, ICSNet~\cite{salazar2024icsnet}, and Mashima et al.~\cite{mashima2023towards} provide simple power grid simulations but are focused on the emulation of the device more than on the physical process, which usually result in static measurements.
Pandapower~\cite{thurner2018pandapower} is a Python-based open-source library designed to enable electric power systems modeling, simulation, and analysis. Key components of the power grid, like buses, power lines, transformers, generators, and loads, are represented as data tables, enabling easy manipulation and customization of network parameters. It provides an element-based modeling approach that allows us to define an electrical network by using nameplate parameters. 
Mosaik~\cite{schutte2011mosaik} is another popular open-source simulation framework designed for modular and distributed simulations of electrical systems. 
It enables the coupling and orchestration of diverse simulators, allowing for the coordinated execution and data exchange. 
Mosaik provides a modular architecture with a central scheduler synchronizing simulators while supporting various APIs and adapters, including Python and Java.

\paragraph{Noise in ICSs}

In a real system, physical process measurements are subjected to noise, an unwanted random variation generated by the system implementation itself, or from interferences with external factors. 
Different forms of noise impact both the physical process~\cite{dehra2018characterization} and the sensors employed to collect the measurements~\cite{ahmed2018noise, ahmed2017hardware}.
Examples of noise sources include thermal fluctuations, electromagnetic interference, mechanical vibrations, and intrinsic limitations of electronic components or sensors, and are also influenced by the generation source~\cite{dehra2018characterization}. 

Measurements of electric voltage could be usually characterized as white noise~\cite{witt2005investigations}, even though some research works proposed other noise models~\cite{vcubonovic2024impact}. 
Different solutions have been proposed to filter noise~\cite{ma2019kalman} or reduce its impact in the real system~\cite{lamo2023impact}. Such forms of errors and variations should be integrated into \ac{ics} simulations to reduce the probability of being detected as a decoy system.

\section{System and Attacker Model}\label{sec:threat}

\paragraph{System Model.}

We assume a scenario where a developer (defender) builds a high-interaction honeypot for an \ac{ics}. The developer is challenged to create a realistic physical process simulation for the honeypot.
The developer does not want to employ a complex model based on differential equations for the \ac{ics} physical process as it is unreliable. This is common in complex dynamic systems such as power grids, where many events can impact the process and cannot be reliably described by a system of equations~\cite{machowski2020power}.

The developer can access a timeseries of physical process measurements from the \ac{ics} \ac{plc} or \ac{hmi}. Such timeseries could be obtained from one-time access to the plant or by employing a public measurement dataset. 
However, the developer is not required to access the \ac{ics} plant after the data collection, which can happen at different times. 

The developer creates a simulation of the real system in its entirety or replicates a small part of it. Small architectural differences are acceptable since the final target does not aim at having a replica of the system but a simulation to be employed in a high-interaction honeypot. %
Moreover, an attacker without access to the real system cannot employ this kind of imperfections to detect the simulations. The developer may start from digital twins~\cite{melesse2020digital} or simulators in the literature~\cite{antonioli2015minicps,salazar2024icsnet,lopez2020honeyplc,lucchese2023honeyics,franco2021survey} and adapt them to resemble the target system. 
\looseness = -1

To prevent identification, the developer follows honeypot anti-fingerprinting best practices, such as the ones proposed by Tay et al.~\cite{tay2023taxonomy}. They ensure that the honeypot network resembles the real one. For example, the attacker can identify industrial devices with valid MAC addresses and see network packets utilizing industrial protocols. 

However, the developer is conscious that an attacker may try to detect the simulation using advanced techniques that investigate the physical process's consistency. Therefore, s/he takes additional design steps to enhance the fidelity of the physical process by adding an appropriate form of noise to the simulated plant.

\begin{figure}[tb]
    \centering
    \includegraphics[width=0.9\linewidth,trim={0 1em 0 1em}, clip]{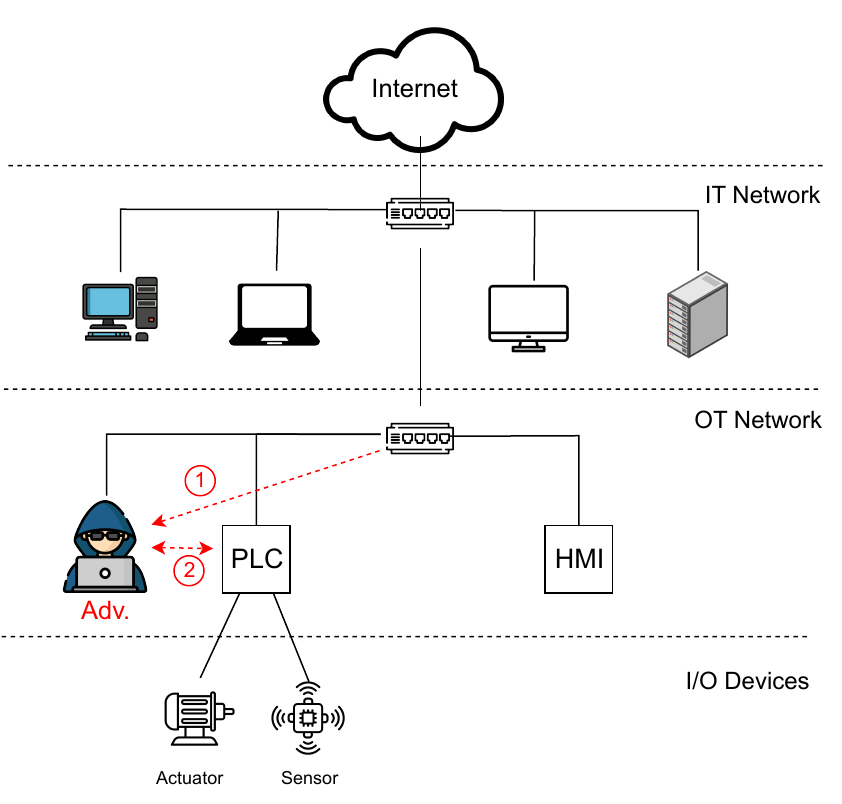}
    \caption{Threat model. \ding{172} represents passive collection of data, while \ding{173} requires the attacker to request values to a \ac{plc}.}\label{fig:threat_model}
\end{figure}

\paragraph{Attacker Model.}
We assume a \emph{remote} attacker with network access to the \ac{ot} network of an \ac{ics} honeypot. As shown in Figure~\ref{fig:threat_model}, the attacker can \emph{passively} collect data or \emph{actively} interact with the honeypot devices, including its virtual \ac{plc} and \ac{hmi} and collect sensor reading from the virtual physical process.

The attacker, who cannot access data from the real system, gets sensor measurements from the defender honeypot as timeseries that contain various noises, including the inherent system noise, the sensor measurement noise, and external noise. The attacker uses such timeseries as a data source to fingerprint the honeypot and categorizes it as a decoy or a real system. 

This is a realistic threat model since many \ac{ics} devices are exposed on the Internet and exhibit none to low-security measures~\cite{mirian2016internet,barbieri2021assessing}. Moreover, industrial honeypots generally employ few to no defenses in the external perimeter to get more interactions from attackers~\cite{franco2021survey}.

\section{\newframework{}}\label{sec:metodology}

We present the design (Section~\ref{subsec:design}) and the implementation (Section~\ref{subsec:implementation}) of \newframework{}, a novel framework that allows identifying the simulations with the best fidelity to replicate a real-world physical process. Our system do not require any differential equation, but only a timeseries of measurements from the real process.  
\looseness = -1

\begin{figure*}[tb]
    \centering
    \includegraphics[width=.9\linewidth]{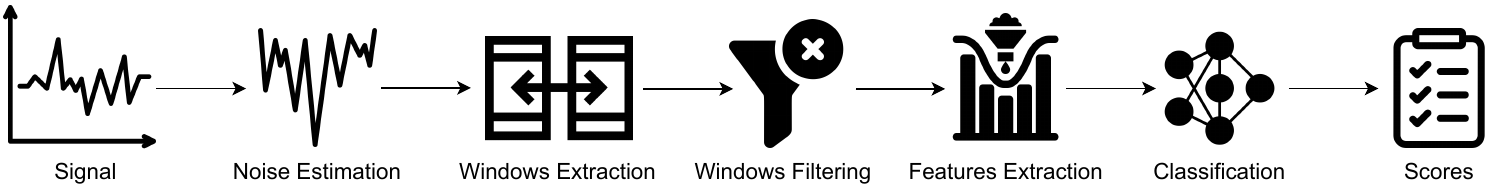}
    \caption{\newframework{} seven-stage pipeline.}
    \label{fig:architecture}
\end{figure*}

\subsection{Design}\label{subsec:design}

As shown in Figure~\ref{fig:architecture}, \newframework{} has a 
\emph{seven-stage pipeline}. The raw signals are collected (Section~\ref{subsec:signal}) and noise estimation is computed from each timeseries (Section~\ref{subsec:noise_estimation}). Then, a windowing approach is taken to prepare the data for classification (Section~\ref{subsec:window_extraction}) with an integrated filtering mechanism to maintain only relevant windows (Section~\ref{subsec:window_filtering}). Then, features are extracted from each window (Section~\ref{subsec:features-extraction}), which are used in the classification phase (Section~\ref{subsec:classification}) which is used to generate the final scores (Section~\ref{subsec:scores}). Next, we describe each pipeline stage.

\subsubsection{Signal}\label{subsec:signal}

The framework collects timeseries from the real physical process and the simulated one the developer wants to benchmark. The timeseries contains measurements from real and virtual sensors connected to real or virtual \acp{plc} or \acp{hmi}. 
Data are collected using a sampling rate of $1$ Hz (i.e., one sample per second). 

A timeseries is defined as: $\hat{x}(t) = x(t) + n(t)$, for a certain period of time $t \in (t_0, t_n)$.
$x(t)$ represents the values that, in a theoretical scenario, will be obtained by sampling the process at a specific time $t$, and it follows the equations that regulate the system. However, real physical processes are subject to noises, modeled by $n(t)$, that sum with $x(t)$. 
In our scenario, only $\hat{x}(t)$ is available because the honeypot developer neither knows the equations that regulate the system $x(t)$, nor the noise $n(t)$, and it is the only signal used as input for \newframework{}.

\subsubsection{Noise Estimation}\label{subsec:noise_estimation}
\newframework{} is based on a profilation of the noise, but due to the absence of ground truth theoretical process $x(t)$, it relies on a noise estimation $\tilde{n}(t)$. Since the developer has some high-level knowledge about the target physical process s/he is replicating (e.g., s/he knows whether the process is a power system, a water treatment, or a manufacturing process), s/he can choose the most suited filter $f( \cdot )$ to be applied on the collected noisy signal $\hat{x}(t)$ to estimate the signal without noise. 
Then, \newframework{} estimate the real noise as the difference between the noisy and the filtered signal, as follows: 
\begin{equation}\label{eq:noise_estimation}
    \tilde{n}(t) = \hat{x}(t) - f(\hat{x}(t)).
\end{equation}

\subsubsection{Windows Extraction}\label{subsec:window_extraction}

At this point, each timeseries $\hat{x}(t)$ is paired with its noisy estimation $\tilde{n}(t)$.
Then, the preprocessing logic applies a sliding window to split the original signal $\hat{x}(t)$ into chunks of size $N$ that overlap by 80\% with each other.
Dividing signals into windows allows for managing signals with different lengths and prevents the dynamics of the system from being preponderant in the classification. 

\subsubsection{Windows Filtering}\label{subsec:window_filtering}
Next, \newframework{} prunes the generated windows by removing the ones containing peaks and big variations in the signal. This allows for more precise extraction of characterizing features, limiting external influences from automated (e.g., a valve opening because a tank is full) or manual (e.g., a user opening an electric switch to turn on a machine) actions on the physical process. This is an essential step to ensure that only real noise is extracted from the original signal and that events happening to the system are not considered.   

The framework performs the pruning by considering the windows $W$ containing values such that: %
\begin{equation}\label{eq:pruning}
    W = \{w_1, \dots, w_N\}, \quad s.t. \; w_i \in [\mu(1-\epsilon), \mu(1+\epsilon)] \quad \forall i\in[1,n]
\end{equation}%
where $\mu$ is the mean value of the window and $\epsilon\in(0,1)$ represents a tradeoff between the natural variation allowed and the removal of potential outliers. High $\epsilon$ alleviates the filter effect and reduces the amount of deleted data. While a low $\epsilon$ removes many samples containing variations with respect to the mean. 
The correct value of $\epsilon$ should be chosen based on the variability of the physical process and the length of the data collection. 

For every window obtained from the signal $\hat{x}(t)$ which has not been pruned by the filtering in Equation~\ref{eq:pruning}, \newframework{} also collects the corresponding window from the paired noise estimation $\tilde{n}(t)$. 
At the end of this preprocessing, the system ends up with a set of windows containing both the raw signal and its corresponding noise estimation, which are the inputs for the feature extraction phase.
\looseness = -1

\subsubsection{Feature Extraction}\label{subsec:features-extraction}

To feed the classification stage, \newframework{} extract features from each window that are relevant to the variations of the signal due to the presence of the noise, without including features that profile the underlying process. 
This allows the freedom in choosing features extracted from windows belonging to the noise estimation $\tilde{n}(t)$ since the decoupling from the underlying process signal has been done with Equation~\ref{eq:noise_estimation}. 
Instead, when dealing with windows generated from the raw timeseries $\hat{x}(t)$, \newframework{} takes extra care to avoid generating biases related to the physical process in our system.

For instance, the system avoids features such as the mean, which is strictly related to the underlying physical process. Instead, it considers features such as the approximate entropy, which measures the complexity or regularity of a time series, and Lempel-Ziv Complexity, which characterizes variations of the system and, in a short period without significant changes in our window, is a good estimation of the noise.  
Moreover, \newframework{} considers the standard deviation ratio $std\_mean\_ratio$ and the variance ratio $var\_mean\_ratio$, defined as:
\begin{equation*}
    std\_mean\_ratio = (\sigma / \mu),
\end{equation*}
\begin{equation*}
    var\_mean\_ratio = (\sigma^2 / \mu),
\end{equation*} 
where $\sigma$ and $\mu$ represent the window standard deviation and mean, respectively. These features allow the signal to be decoupled from superimposed noise generated by physical components as much as possible. 
\looseness = -1

\subsubsection{Classification}\label{subsec:classification}

The features extracted in this pipeline are then fed to a \ac{ml} model trained to differentiate between real physical processes and simulations. The task can be formulated as a binary classification problem, where data are labeled as \textit{real} or \textit{simulated} based on their origin.  
The models take into account a feature set related to a single window and output a prediction. Various \ac{ml} models are suitable for the purpose and may be chosen based on recall scores but also on the final model complexity~\cite{bernieri2019evaluation}. %

\subsubsection{Scores}\label{subsec:scores}
Having a good-performing model is essential to ensure a solid benchmarking process. However, in this case, the framework aims at maximizing the true positives (i.e., real samples correctly classified as real) possibly at the cost of misclassifying some fake system. When this happens it is a suggestion that such a sample is similar enough to the real process that it can be passed as a real system. To enhance this effect, our system is designed to output \textit{the probability for each sample to be classified as real} instead of thresholding the output model probability and directly predicting the class. This allows for the generation of a \textit{fidelity score} for every window provided to the model. The final score for each data source (and, therefore, on each noise) is computed by averaging the probabilities on every window associated with it.

\subsection{Implementation}\label{subsec:implementation}

We implemented \newframework{} using Python. 
Datasets are saved as CSV files to be easily imported into the code. 
Noise is estimated by employing a filter that can be easily changed to best suit each scenario. 
Then, the windowing approach is applied to extract windows of a predefined size. Every window is checked on a simple implementation of Equation~\ref{eq:pruning} and windows not satisfying the equation are dropped, taking care of keeping the sync between different measurements.

To extract relevant features from the windows, we employed the \texttt{tsfresh} Python package~\cite{tsfresh}. The package automatically computes statistical and mathematical characteristics from time series data and selects the most relevant ones for regression or classification tasks.
The extracted features include, among others, statistical measures (e.g., mean, variance, skewness, kurtosis), frequency-domain features (e.g., Fourier coefficients, spectral entropy), and time-series properties (e.g., autocorrelation, trend strength, peaks, crossings).

We let \texttt{tsfresh} identify the most suitable features for the noise estimations $\tilde{n}(t)$, selecting a set of features we employed for all subsequent extractions. %
Instead, we manually limit the possible features extracted from the noisy signal $\hat{x}(t)$. We carefully choose 9 features (described in Appendix~\ref{app:features}) that profile the signal variations due to the noise without taking into account the underlying signal, in addition to the two new features introduced in Section~\ref{subsec:features-extraction}.

Finally, we employed Scikit learn~\cite{pedregosa2011scikit} to perform the classification step. It is a well-known Python implementation of different \ac{ml} models and support functions. Moreover, it allows for the extraction of class probabilities instead of classification outputs from the models. We employed this strategy to compute the final score of our benchmarking systems. 

\section{\newframework{} at work: the EPIC case study}\label{sec:case}

We evaluated \newframework{} on EPIC~\cite{adepu2019epic}, a state-of-the-art power grid testbed. We chose a power grid because it is widely employed in security research in \ac{ics}~\cite{PowerGrid}. Moreover, we also needed ground truth data from a real plant; this requirement reduced the possible scenario to consider since datasets coming from real \ac{ics} are still scarce in the literature~\cite{conti2021survey}. We note that our framework can also be used in other \ac{ics} domains.

For our case study, we use the methodology and seven-stage pipeline introduced in Figure~\ref{fig:architecture}. Section~\ref{subsec:new_dataset} describes the dataset we use to test \newframework{} (first pipeline stage). Then, Section~\ref{subsec:new_evaluation} shows some insight into the fine-tuning of \newframework{} to provide the best results on this scenario (second to sixth pipeline stages). Next, Section~\ref{subsec:new_results} tests the overall framework and discusses the results (last pipeline stage).

\subsection{Dataset Generation}\label{subsec:new_dataset}

This section describes the EPIC ground-truth dataset (Section~\ref{subsec:datasets}), how we replicate in different simulators a portion of the real testbed (Section~\ref{subsec:simulations}), and how we augmented the fidelity by adding noise to our simulations (Section~\ref{subsec:additional_noise}). A summary of the different timeseries that compose our dataset is summarized in Table~\ref{tab:data-summary}. The dataset is then used as input signals in our pipeline.

\begin{table}[tb]
    \caption{Summary of the employed datasets. Each noise has been applied to both simulators (Sim.), namely Pandapower and Mosaik.}
    \centering
    \resizebox{.9\columnwidth}{!}{%
    \begin{tabular}{@{} lccl @{}}
    \toprule
\textbf{Source} & \textbf{Real} & \textbf{Noise} & \textbf{Parameters} \\ \midrule
EPIC~\cite{adepu2019epic}            & \yes          & \yes    & -         \\
Plain Simulation          & \no           & \no                     & -   \\
Sim. + uniform  & \no           & \yes                &  $\sigma_{u}$=0.01  \\
Sim. + gaussian1  & \no           & \yes            & $\sigma_{g}$=0.01 \\
Sim. + gaussian2  & \no           & \yes            & $\sigma_{g}$=0.05 \\
Sim. + poisson  & \no           & \yes            & $\sigma_{p}$=0.01, $\lambda_{p}$=1.5 \\
Sim. + laplace  & \no           & \yes            & $\sigma_{l}$=0.01 \\
Sim. + pink  & \no           & \yes            & $\sigma_{p}$=0.01 \\
Sim. + GMM  & \no           & \yes            & $\sigma_{g}$=0.02 \\
Sim. + gaussian + uniform  & \no           & \yes            & $\sigma_{g}$=0.01, $\sigma_{u}$=0.01 \\
Sim. + laplace + uniform  & \no           & \yes            & $\sigma_{l}$=0.01, $\sigma_{u}$=0.01 \\
Sim. + laplace + poisson  & \no           & \yes            & $\sigma_{l}$=0.01, $\sigma_{u}$=0.01, $\lambda_{p}$=1.5 \\
Sim. + VRAE  & \no           & \yes           & $input\_weight=0.99$  \\

\bottomrule     
\end{tabular}%
}
    \label{tab:data-summary}
\end{table}

\subsubsection{Ground-Truth Datasets}\label{subsec:datasets}

\begin{figure}[tb]
    \centering
    \includegraphics[width=.9\linewidth,trim={0 0 0 0},clip]{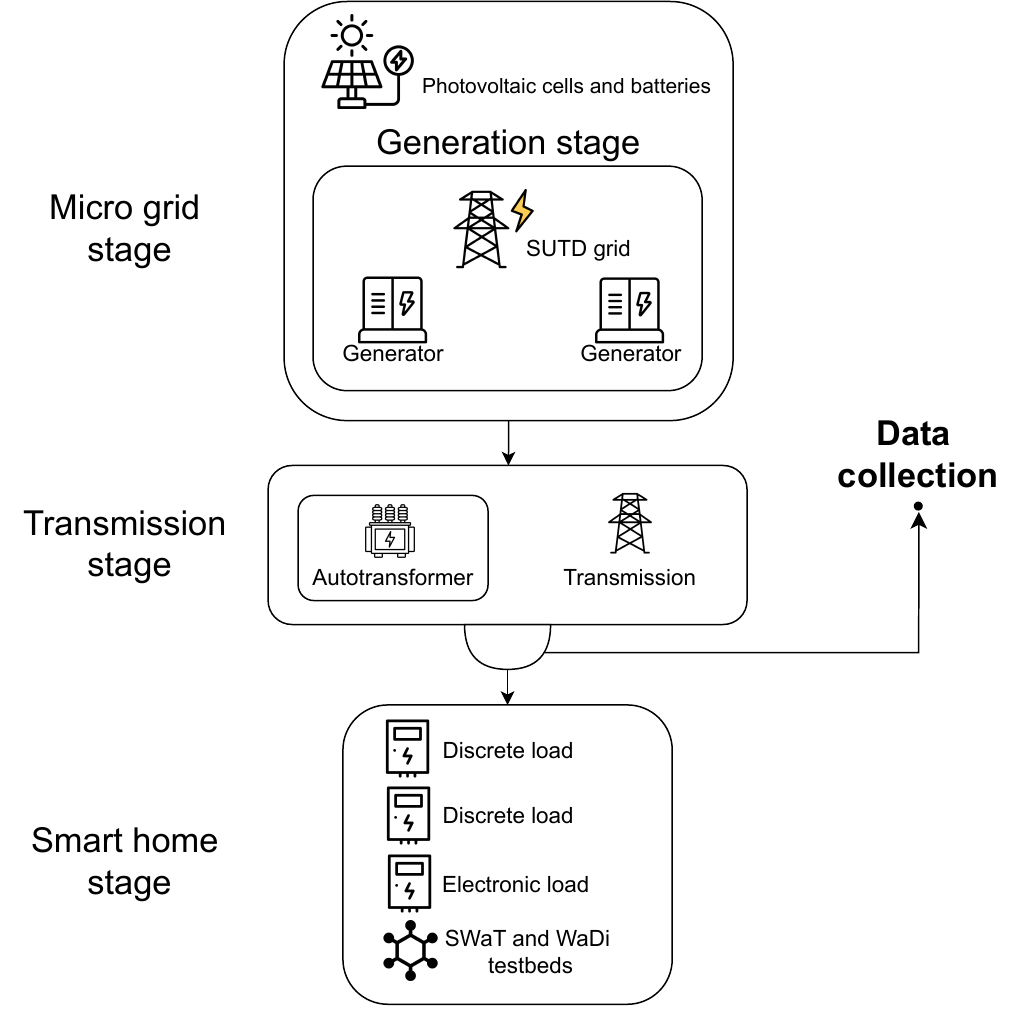}
    \caption{EPIC testbed visual description.}
    \label{fig:epic_architecture}
\end{figure}

We needed a dataset containing physical process measurements from a real-world industrial plant or testbed. We found an adequate dataset from a paper on EPIC presented in 2019~\cite{adepu2019epic}. The EPIC datasets include eight scenarios in which the system runs under different conditions, such as different numbers of active generators and loads. The dataset time series are recorded from sensors and actuators, capturing 30 minutes of operation per scenario. For each case, network traffic data is preserved in packet capture files, while system readings are stored in CSV files, offering a view of the physical process behavior. 

EPIC is a sophisticated platform designed to emulate real-world power grid operations. Developed by iTrust, the testbed includes four interconnected stages, as summarized in Figure~\ref{fig:epic_architecture}: Generation, Transmission, Microgrid, and Smart Home, providing a comprehensive environment for studying smart grid dynamics~\cite{10.1145/3198458.3198461}.   

EPIC enables testing of various operational conditions, such as synchronization of generators, integration of renewable energy sources, and dynamic load management. It utilizes \acp{plc} for control and supervision, while a SCADA system ensures centralized monitoring. This characterizes flexibility, making it a realistic system for analysis and comparison. Among the many case studies, EPIC has been used to study the impacts of power supply interruption and cyberattacks~\cite{kandasamy2019investigation}.

Since the EPIC dataset contains a small number of samples with respect to the ones we can generate with the simulators, we decided to increase, when needed, the number of samples by performing a data augmentation preprocessing on the final generated features by using SMOTE~\cite{chawla2002smote}. %
In addition to increasing the real sample size in the dataset, this step helps prevent overfitting and enhance the generalizability of our model.  
However, we employed these samples for training, keeping the testing phase clean from these artificial samples.

\subsubsection{Simulations}\label{subsec:simulations}

We decided to simulate the underlying physical process of the EPIC testbed using two popular tools for power grid simulations, Pandapower~\cite{thurner2018pandapower} and Mosaik~\cite{schutte2011mosaik}. Both tools rely on rule-based and physics-based modeling rather than data-driven or statistical approaches. This ensures the generated data reflects realistic operational behavior and allows an attacker to interact with the system.
We extracted information on its architecture (e.g., number of generators) from documentation describing the system~\cite{adepu2019epic, tan2024high}. Since the available references do not mention the base values employed in EPIC, we extracted a rough estimation from the EPIC dataset. In particular, we set a base current of $20A$, a base voltage of $240V$, and a base frequency of $50Hz$. We do not need to precisely match EPIC values since \newframework{} focuses on the noise applied on top of the physical process and not on the exact values of the process itself.   
Pandapower and Mosaik do not generate process noise, so the output $ x*(t)$ is a theoretical simulation of $x(t)$.

The data generated from the simulation, to be as coherent as possible with EPIC dataset, contains direct measures of \emph{Voltage} ($V1,V2,V3$), \emph{Current} ($I1,I2,I3$), and \emph{Frequency}. Moreover, we compute the \emph{Reactive power} as $S = \sqrt{3} V  I$ ($VI \in \{ V1,V2,V3 \}$), the \emph{Real power} as $P = S \cos(\phi)$, and the  \emph{Apparent power} as $Q = S \sin(\phi)$, where $\phi = 120^{\circ}$  represents the phase of the EPIC testbed, while $V$ and $I$ represent respectively the average values over the three phases for voltage ($V1, V2, V3$) and for current ($I1, I2, I3$).

\subsubsection{Noise Simulation}\label{subsec:additional_noise}

We introduced different approaches 
for adding various noises $n_i(t)$ to the underlying simulations, to generate a signal $\hat{x}^*(t) = x^*(t) + n_i(t)$ that reproduces as close as possible the genuine signals $\hat{x}(t)$. Then, we use \newframework{} to rank the possible noises $n_i(t)$ and select the ones that make the simulated data look closer to the real $\hat{x}(t)$.

We select the noise candidates by looking at papers discussing noise in power systems. Characterizing noise from a power grid is complicated since many factors can interfere with the desired signal, depending on internal and external conditions, as described in Section~\ref{sec:background}.

We started by adding \textit{normal (Gaussian) noise} to the simulated data, which is known to represent simple random deviations introduced by thermal noise~\cite{lamo2023impact}. 
There is no clear definition of the correct standard deviation value for such a noise distribution on electric power systems. As discussed in the literature~\cite{brown2016characterizing, tripathy2009divide, shi2012adaptive, zhang2013two, huang2015dynamic}, standard deviations range between 0.002 and 0.15 pu. We choose two values in the middle to model our zero-mean Gaussian noises (0.05 and 0.01). 

However, Gaussian noise alone cannot model an electric power system~\cite{chen2019robust, wang2017assessing, zhou2014capturing}. Actuators, sensors, transmission lines, and external factors generate additional noise that does not follow a Gaussian distribution. 
Following research in the literature~\cite{vcubonovic2024impact, martinez2020analysis, baldick1997implementing} and our intuition, we tested other noise distributions, such as Laplace and uniform noise distribution. 
\looseness = -1 

The Laplace noise represents impulse noise generated by electromagnetic interferences and may represent external noise sources~\cite{coletta2018transient}. 
Uniform noise may be present in power systems, due, for instance, to quantization noise~\cite{baldick1997implementing}.
We also considered other noises generated from Poisson and Pink distributions, which may represent the various noise sources of a real power system. 
Moreover, we employed a \ac{gmm} which can be regarded as a linear combination of $k$ independent Gaussian models and, in theory, can fit any distribution~\cite{vcubonovic2024impact}. We extract their parameters from an unprocessed EPIC trace and use it as an additional noise source.

\begin{figure}[tb]
    \centering
        \includegraphics[width=\columnwidth]{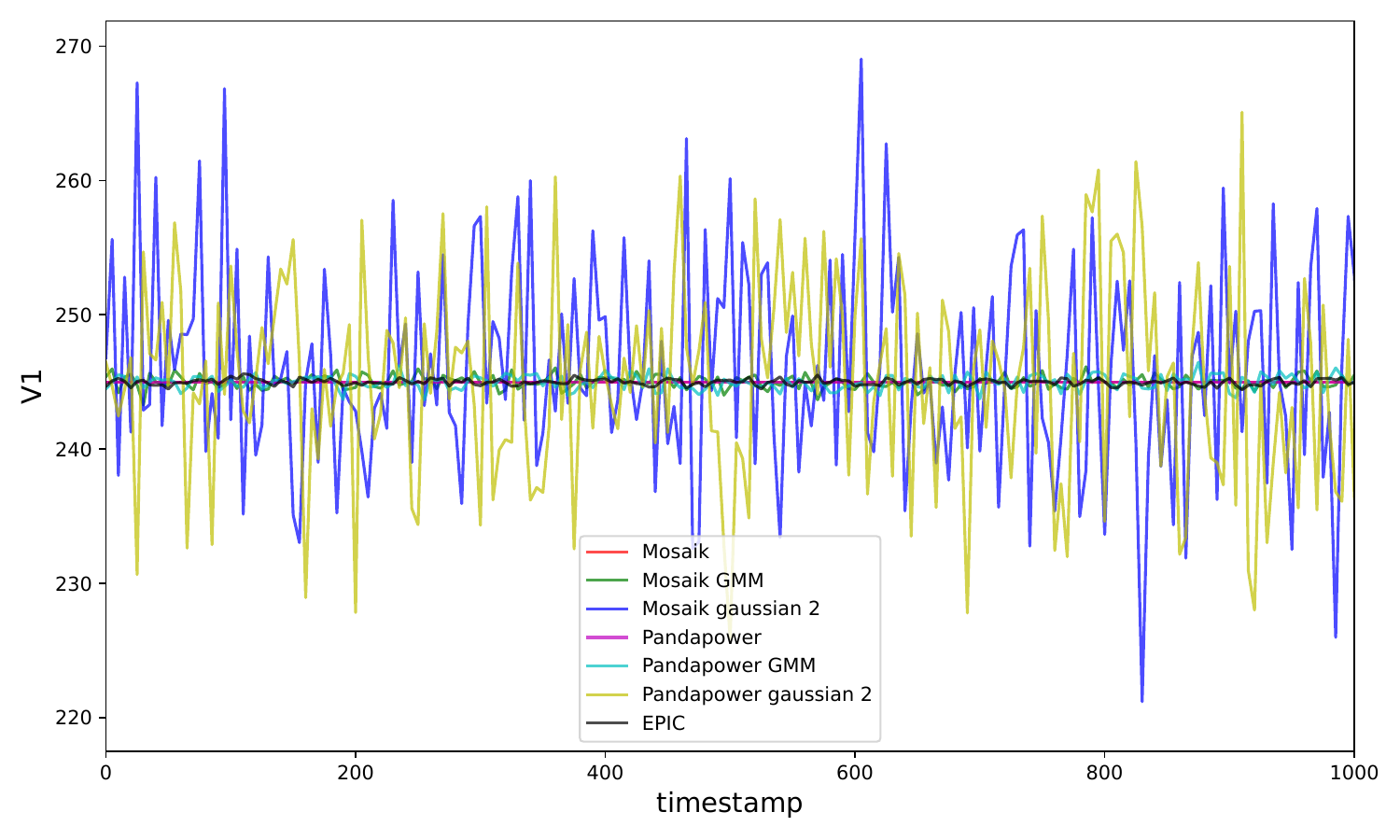}
    \caption{Example of real (EPIC) and simulated (Mosaik, Pandapower) power grid timeseries with some sample noises. Simulations without noise are almost flat and, for the most part, superimposed by others.}
    \label{fig:curves}
\end{figure}

The approaches considered up to now sample the noise errors from the same distribution throughout the process. While this may be the case for certain power grids, external interferences may change over time and based on external factors. 
A dynamic approach involves using generative methods. An autoencoder can be trained on real data and employed to make simulation data closer to the real process in real-time.

We test this strategy by employing a \ac{vrae} thanks to its causality that allows the generating samples temporarily dependent on previous elements in the time series. We trained our \ac{vrae} employing the same architecture of Chung et al.~\cite{chung2015recurrent} on the noise estimation generated as depicted in Section~\ref{subsec:noise_estimation}. During generation, the \ac{vrae} gets as input a clean signal (i.e., without any noise) coming from one of the two simulators.

The output is a new signal resembling the one provided as input, but with additional noise. Since the \ac{vrae} only manages noise, we set the input weight to a high value (0.99) to ensure the output is mainly controlled by the signal in input. Our \ac{vrae} has been trained for 11 epochs. We refer the reader to the original paper for more details on the model architecture~\cite{chung2015recurrent}. 

Figure~\ref{fig:curves} shows an example of different noises applied to our simulation for the voltage V1, together with the real data from the EPIC dataset. %

\subsection{\newframework{} Fine-Tuning}\label{subsec:new_evaluation}

We fine-tune the parameters of \newframework{} on the EPIC power grid employing the dataset we built as discussed in Section~\ref{subsec:new_dataset}.
We analyze the system parameters, such as the balancing ratio between real and fake data in the training set (Section~\ref{subsec:dataset-balancing}), the window length (Section~\ref{subsec:window_length}), the selection of the number of features to be used for classification (Section~\ref{subsec:features-reduction}), and the \ac{ml} model to be used (Section~\ref{subsec:model-selection}).
\looseness = -1

As a filter $f(\cdot)$ to estimate the noise, we employed a Kalman filter~\cite{ma2019kalman}. Based on our tests, we applied two different $\epsilon$ values to the window pruning (Equation~\ref{eq:pruning}): $\epsilon=0.1$ for filtering individual values, $\epsilon=0.3$ for filtering all values simultaneously. This adaptive choice of $\epsilon$ provides a trade-off between filtering capabilities and data retention. 

A lower $\epsilon$ resulted in a more stringent filter, reducing the amount of available data, while a higher $\epsilon$ made the filter more permissive, potentially retaining outlier values. 
The higher $\epsilon$ value was chosen for the case where the filter is applied to all values simultaneously since, in this case, a window is retained only if all values within it satisfy the pruning criterion (Equation \ref{eq:pruning}). Thus, a more permissive threshold was necessary to maintain sufficient data for further processing of values.%

For the evaluation, we employed the recall metric, which is defined as $Recall = {TP}/(TP + FN)$, 
where True Positive (TP) counts real samples correctly identified as real, while False Negative (FN) identifies real samples classified as simulated.
Recall measures the TP rate, indicating the rate of real samples classified as real. 
We perform the experiments on every measure in our dataset, in addition to collecting all the measures together (\emph{allvalues}). For clartity, in most plots we show only results on  some representative measurements.  
\looseness = -1

\subsubsection{Dataset Balancing}\label{subsec:dataset-balancing}

Because the EPIC dataset contains little real data compared to the large amount of data generated from the different simulations, we tried various levels of balancing the dataset.  
We employed 90\% of the real data for training and 10\% for testing. Since we can benefit from a huge dataset of simulations, we extract 30\% of the data for training and 35\% for testing, using stratification to ensure that every noise appears equally. 

Then, we selected different balancing levels and trained a \ac{rf} classifier on sample settings to monitor the performances. When needed, we employed SMOTE~\cite{chawla2002smote} to perform data augmentation on the training data belonging to the real class. 

\begin{figure}[tb]
    \centering
    \includegraphics[width=\columnwidth]{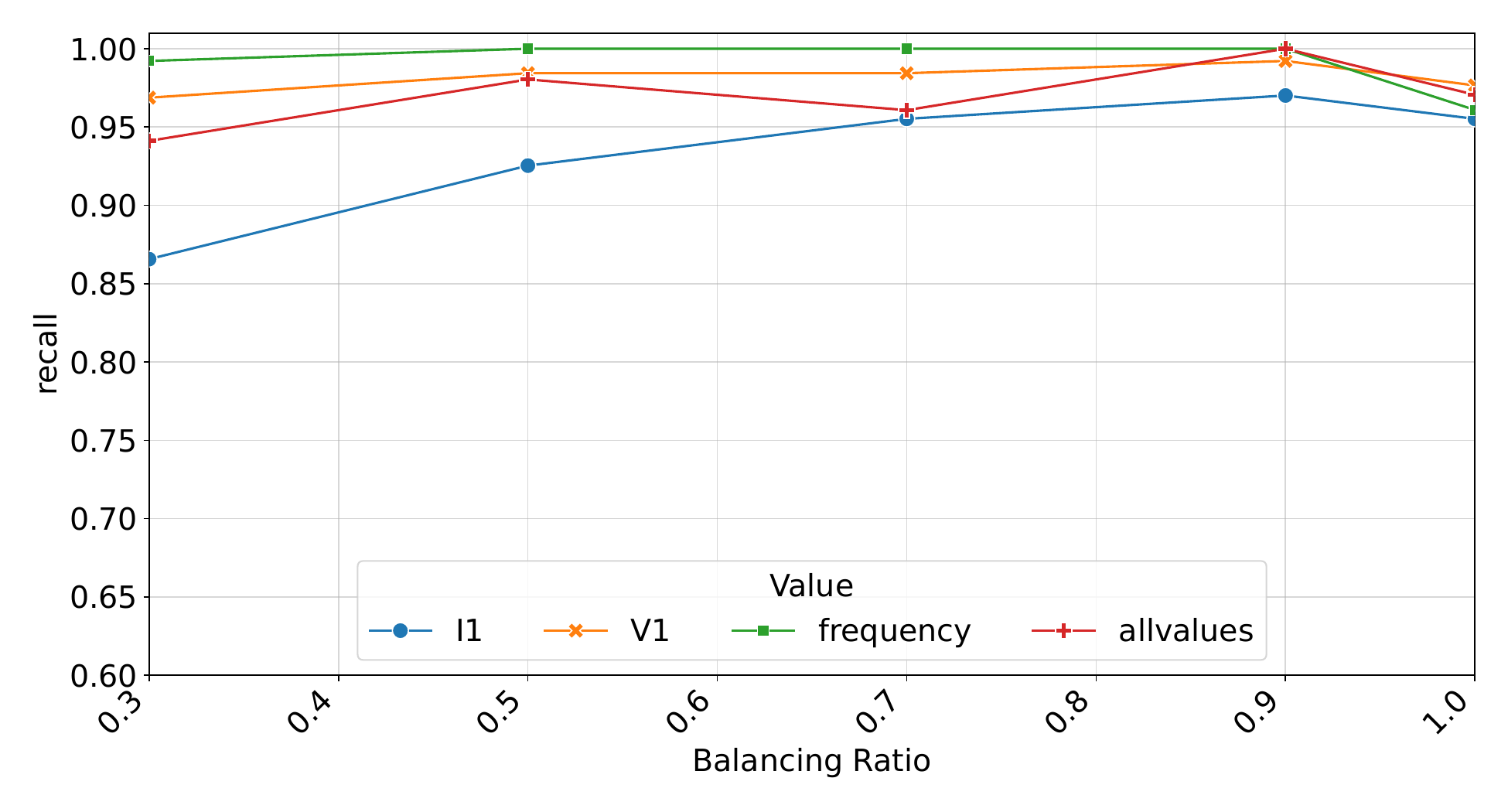}
    \caption{Recall while varying the balance ratio. A higher balancing ratio indicates a more balanced dataset. I1 is the first current phase, while V1 is the first voltage phase.}\label{fig:recall_balRatio}
\end{figure}

We can see from Figure~\ref{fig:recall_balRatio} that the recall increases with the balancing ratio, suggesting, as expected, that increasing the ratio of real samples in the training data allows for a better TPs rate. This is the desired goal of our framework since we want to measure the distance of simulated samples from the real ones that need to be classified correctly as much as possible. 

\subsubsection{Window Length Analysis}\label{subsec:window_length}

As discussed in Section~\ref{subsec:window_extraction}, the window length is the number of samples used to compute the features given to the models as input. Therefore, it represents the minimum duration of the data collection to get the first decision on the physical process under test. Moreover, different window sizes capture different behaviors in the data distributions and provide different samples for training and testing. Each window is classified alone for \newframework{}, and then all the scores are averaged to obtain the final noise ranking.

\begin{figure}[tb]
    \centering
    \includegraphics[width=\linewidth]{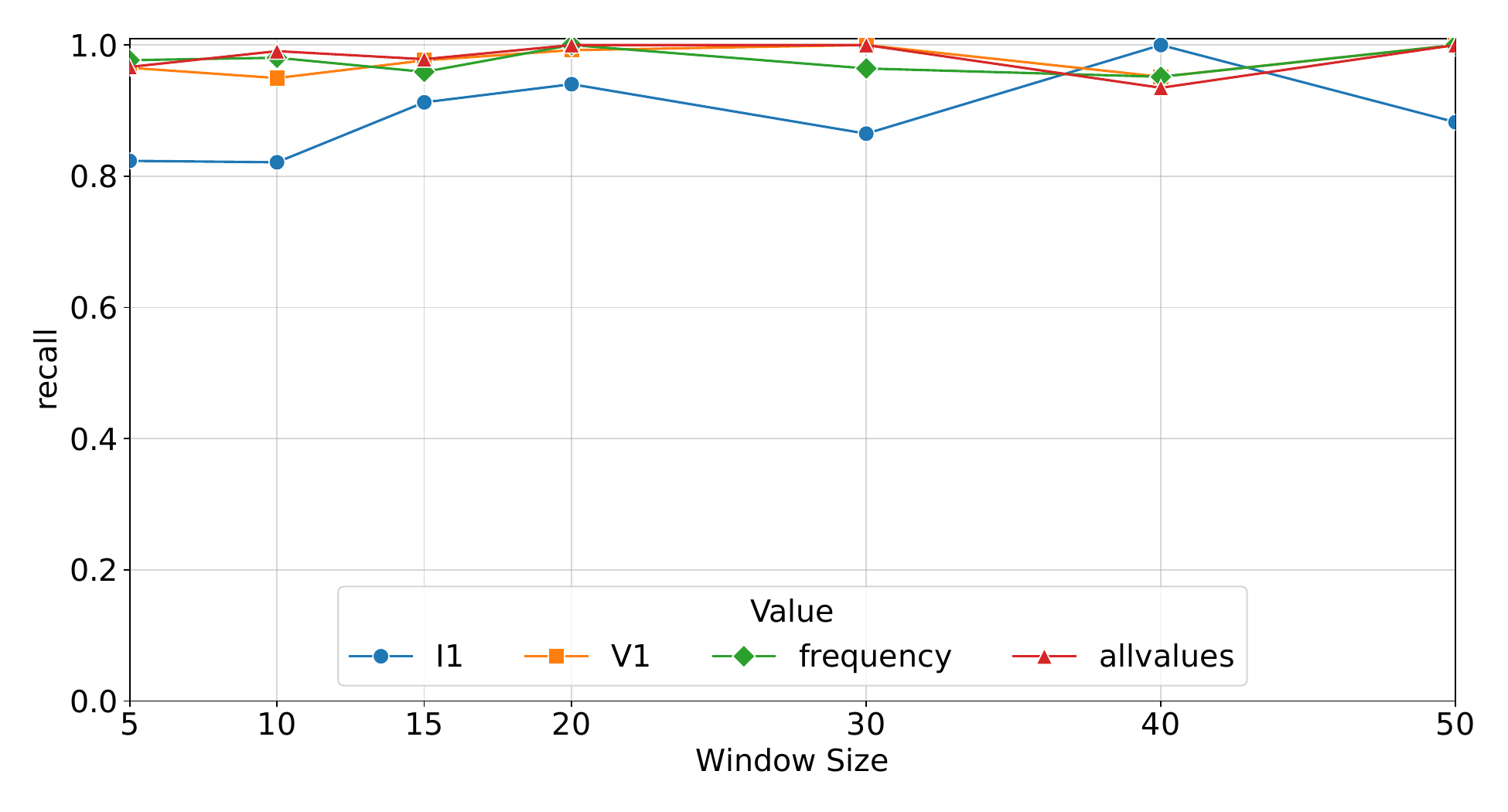}
    \caption{Recall when changing the window size.}
    \label{fig:windows}
\end{figure}

We identify the window length by experimenting on the \ac{rf} classifier with the best results obtained up to now. Results are shown in Figure~\ref{fig:windows}. The recall does not change much while increasing the window size, except for I1, which shows fluctuating values. However, we can see a convergence of high recall for every value considered with a window size of 20, which has been chosen as the default value in our evaluation.

\subsubsection{Feature Reduction}\label{subsec:features-reduction}

\begin{figure}[tb]
    \centering
    \includegraphics[width=\linewidth]{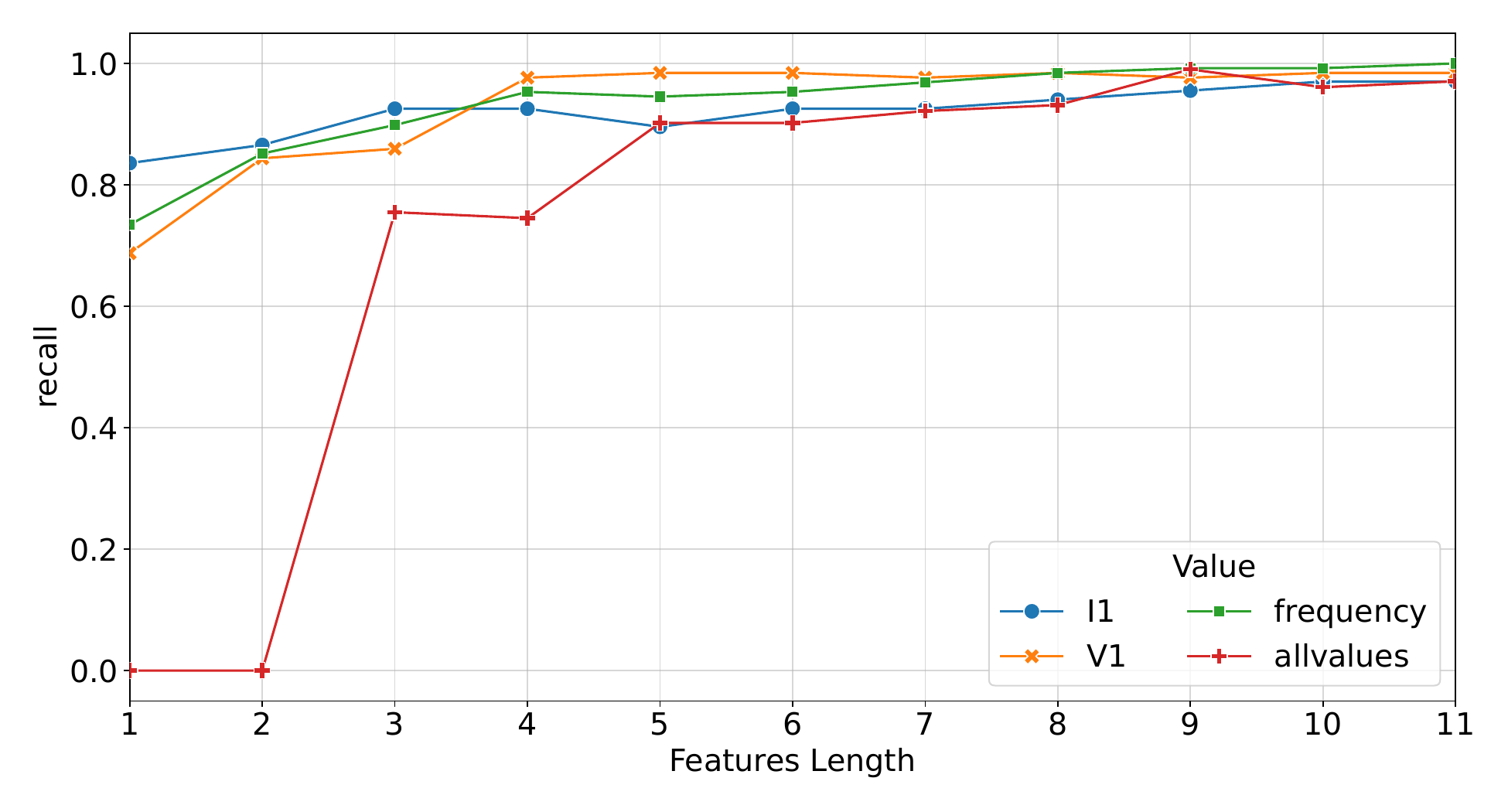}
    \caption{Recall when adding one feature at a time ordered by importance. %
    }
    \label{fig:features_f1}
\end{figure}

Based on the most features classification provided by \texttt{tsfresh}~\cite{tsfresh}, we tested their importance by training the model adding one feature at a time, from the most important to the least important, up to 15 features. We tested it for V1, I1, frequency, and all the values together. Results are shown in Figure~\ref{fig:features_f1}.
The first few features influence the measures, while additional features only refine the classification.  We see an increase in the recall score, which is particularly pronounced in the first five features. However, the first peak for all the tested features is reached with eleven features, which represents the best number of features for our study.

\subsubsection{Model Selection}\label{subsec:model-selection}

Next in our case study, we merged the results of all the above experiments to identify the most suited model for our task. We formulate the task as a binary classification problem, where training data comes from real and simulated systems. The model is asked to identify between real and simulated data.

We compare different lightweight \ac{ml} models such as \ac{dt}, \ac{lr}, \ac{knn}, \ac{rf} and \ac{ab}. We also employ a deep learning approach with a simple \ac{nn}. For consistency, the \ac{nn} has been trained on the same set of features. For every model, we fine-tuned the hyperparameters using grid search to maximize the recall.

\begin{figure}[tb]
        \centering
        \includegraphics[width=\columnwidth]{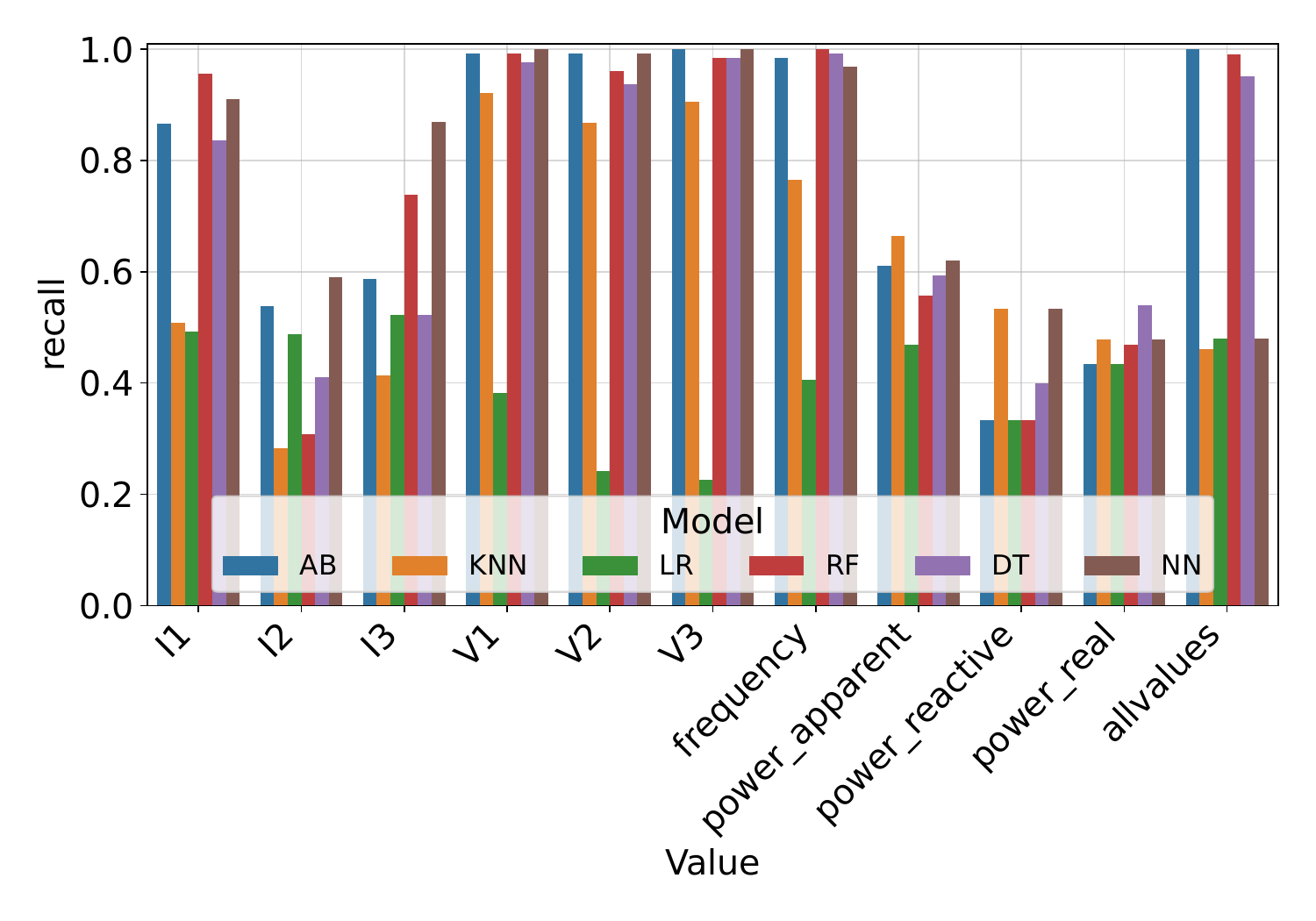}
        \caption{Recall employing different models.}
        \label{fig:binary_scores}
\end{figure}

Figure~\ref{fig:binary_scores} summarizes the performances of the various models. As explained in Section~\ref{subsec:classification}, we want to reach high recall to ensure a good representation of all the EPIC samples as real. 
We can see how different models performed quite well, except \ac{lr}, which failed to identify most of the EPIC sample as real. 
\ac{ab}, \ac{nn}, \ac{dt}, and \ac{rf} are the best-performing models, with some differences. For instance, the \ac{nn} performs well in the current I3, while \ac{rf} is the best model for the first phase of the current I1. We prefer \ac{ab} and \ac{rf} to deep architectures such as the \ac{nn} that may lead to overfitting, but all of them could be considered as good candidate models for \newframework{}.

\subsection{Results}\label{subsec:new_results}

This section presents the results of our EPIC case study. We show the capabilities of \newframework{} to provide a fidelity score value to each simulation and for each measure and discuss its strengths and limitations. We analyze the relevance of the employed features in Section~\ref{subsec:features-analysis}, the fidelity scores in Section~\ref{subsec:eval_fidelity_score}, and discuss the applicability in a dynamic system in Section~\ref{subsec:eval_delta}. 

\subsubsection{Features Analysis}\label{subsec:features-analysis}

\begin{figure}[tb]
    \centering
    \includegraphics[width=\columnwidth]{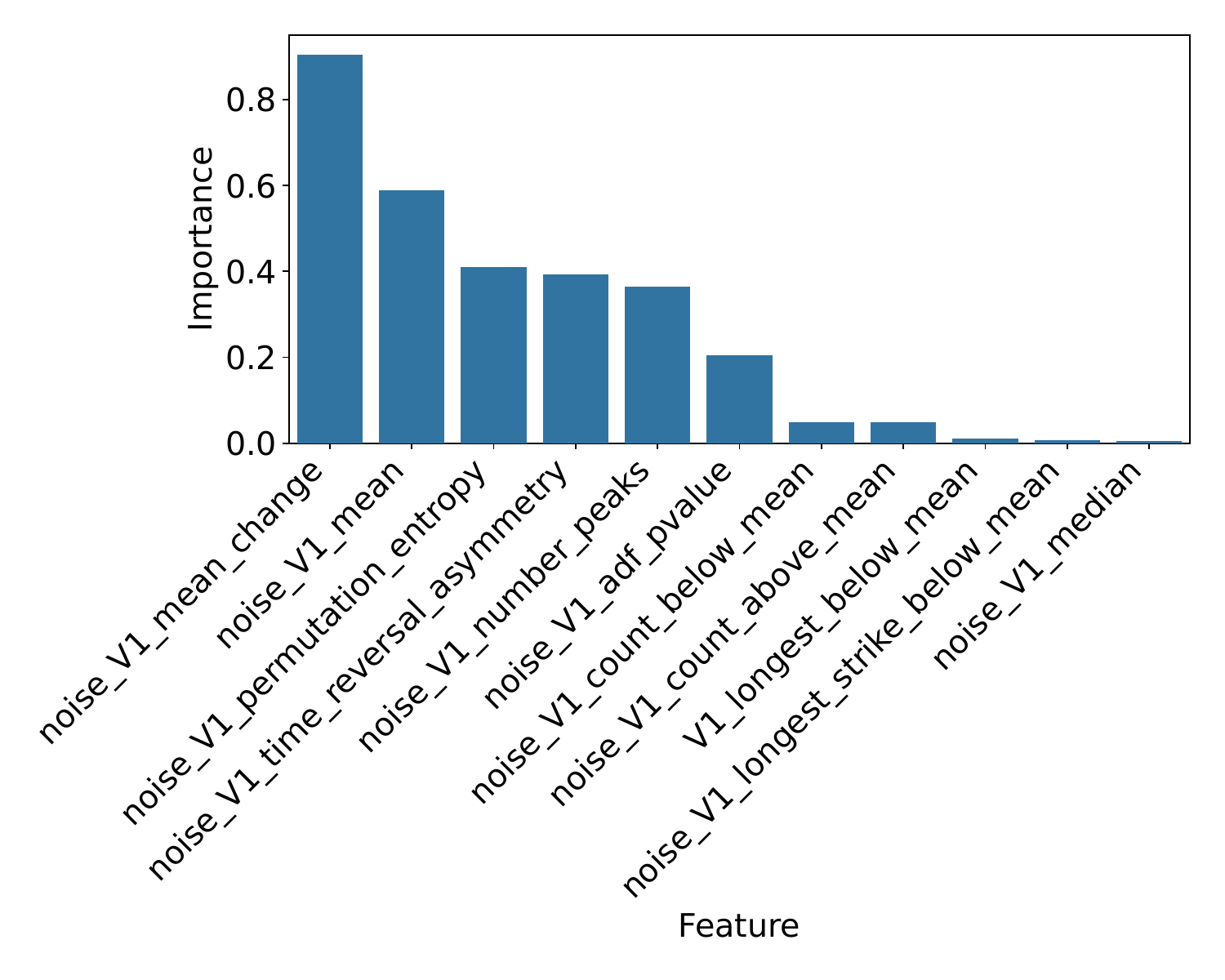}
    \caption{Most important features for V1. Features beginning with ``noise'' point to features extracted from the noise estimation.}
    \label{fig:feature_list}
\end{figure}

Understanding which are the features that impact the most in the classification is essential to comprehend the capabilities of the system.
Figure~\ref{fig:feature_list} shows the most important features we isolate based on %
p-values for the voltage V1 case. We can see how the noise mean (i.e., the mean of the estimated noise within each window) is the most important feature, followed by the permutation entropy of the noise. Five more features extracted from the noise estimate follow suggesting that the noise estimation is practical and its features are meaningful for the classification. Then, the longest above mean feature extracted from the original signal suggests that, even to a smaller extent, meaningful information is extracted from the original signal as well. 
Similar graphs for the other measurements are available in our GitHub repository.

\subsubsection{Simulations Fidelity Scores}\label{subsec:eval_fidelity_score}

\begin{figure}[tb]
    \begin{subfigure}{0.48\textwidth}
        \centering
        \includegraphics[width=\linewidth]{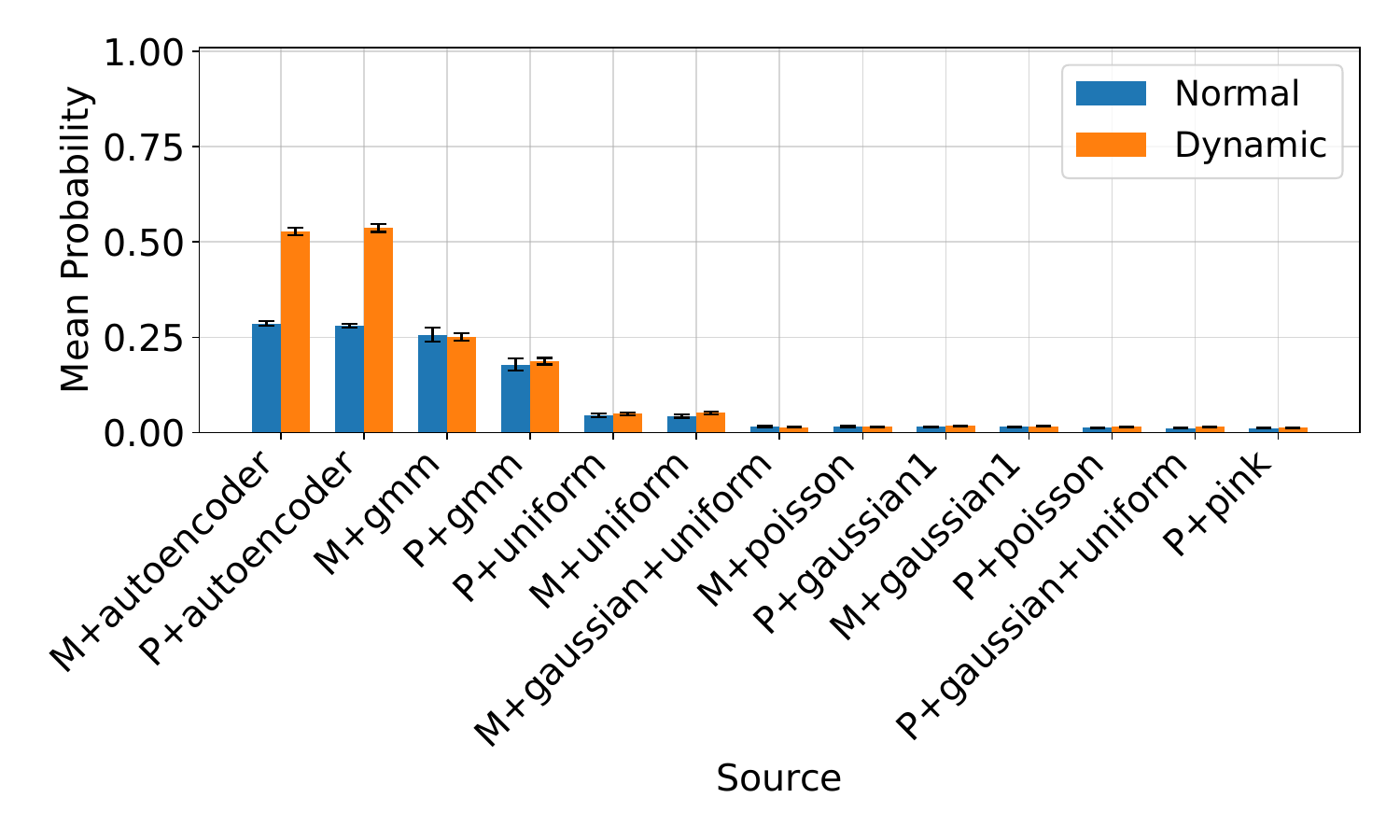}
        \caption{Voltage V1.}
        \label{subfig:sim_scores_V1}
    \end{subfigure}

    \begin{subfigure}{0.48\textwidth}
        \includegraphics[width=\linewidth]{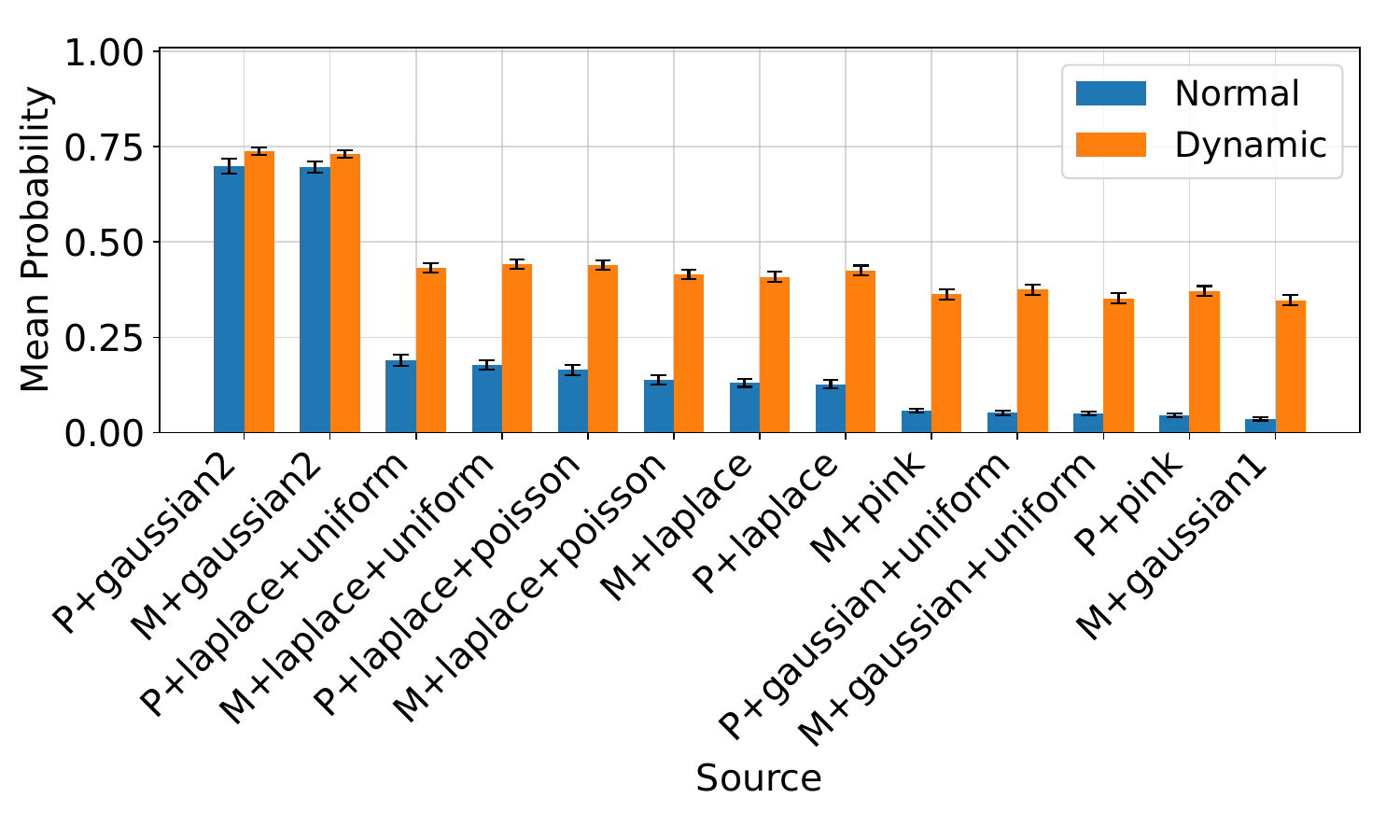}
        \caption{Current I1.}
        \label{subfig:sim_scores_I1}
    \end{subfigure}

    \caption{Resemblance of a real system of each simulation (M: Mosaik, P: Pandapower) with different noises applied. Blue scores represent normal samples (used for training) discussed in Section~\ref{subsec:eval_fidelity_score}, while orange represents the dynamic scenario (not used in training) discussed in Subsection~\ref{subsec:eval_delta}. Error bars indicates the standard error.}
    \label{fig:sim_scores}
\end{figure}

Figure~\ref{fig:sim_scores} shows each window's mean probability of being classified as a real sample (blue bars). High values represent noise that may easily resemble the real data, while low values indicate high chances for the system to recognize the samples as a simulation. 
In particular, Figure~\ref{subfig:sim_scores_V1} shows the probabilities related to the first voltage phase V1. The graph suggested that \ac{gmm} is the most well-performing noise for these kinds of measurements with a score of 0.267, even if the autoencoder noise follows with almost the same score (0.265). 

Figure~\ref{subfig:sim_scores_I1} represents a different situation for the first current phase I1. Here we can see a high peak of the Gaussian noise with high standard deviation, representing the optimal noise for current values. Then, it is interesting to see that uniform noise follows obtaining good results, and it may indicate errors introduced by the analog-to-digital converters in the \ac{plc}. 

We cannot plot the graphs for all the measurements for space reasons (the reader will find it in the GitHub repository), but Table~\ref{tab:best_noises_summary} summarizes the best noise for each value, indicating the probability for such a sample to be categorized as real.
We can see how similar measures share the same noise. The Gaussian distribution well represents current noise, while the voltages are well represented by \ac{gmm} and autoencoders. In particular, similarly to Figure~\ref{subfig:sim_scores_V1}, V2 and V3 show scores for \ac{gmm} that are close to the autoencoder.

Our autoencoder also represents frequency noise, while powers follow Gaussian noises, which are probably derived from the current measurements. 
Interestingly, the power reactive follows a Laplace + Uniform noise, even though other noises such as Poisson and Uniform reach similar probabilities. When considering all values, the \ac{gmm} is the best distribution to replicate the noise. 

Similar measures (e.g., current) share similar noise distributions (e.g., Gaussian), which is at least partially related to the type of sensors employed to collect the measurements. 
While it is possible to identify the swapping of a sensor with another identical replica through a difference in the generated noise~\cite{ahmed2018noise, ahmed2017hardware}, the difference magnitude is more significant when considering different categories of sensors. Still, similar patterns may be observed for probes measuring the same quantity. 
These considerations suggest that the better strategy for a developer is to take a selective approach and employ different noises for each signal s/he needs to simulate.

\begin{table}[tb]
    \caption{Summary of the best noises for every measurement. Prob. Indicates the associated probability of misclassification as real, while Delta is the difference in probability in the case of dynamic samples.}
    \centering
    \begin{tabular}{@{} llcc @{}} 
    \toprule
\textbf{Value}  & \textbf{Best Noise} & \textbf{Prob.} & \textbf{Delta} \\ \midrule
voltage V1              & GMM                 & 0.267   &  0.064     \\
voltage V2              & Autoencoder         & 0.397   &  0.078      \\
voltage V3              & Autoencoder         & 0.356   &  0.076       \\
current I1              & Gaussian2           & 0.688   &  0.052      \\
current I2              & Gaussian2           & 0.365   &  0.074      \\
current I3              & Gaussian2           & 0.525   &  0.065      \\
frequency       & GMM                 & 0.331   &  0.233       \\
power\_apparent & Gaussian2           & 0.605   &  0.039       \\
power\_reactive & Laplace+Uniform     & 0.133   &  0.045      \\
power\_real     & Gaussian2           & 0.707   &  0.073     \\
all values      & GMM                 & 0.418   &  0.119     \\
\bottomrule
\end{tabular}
    \label{tab:best_noises_summary}
\end{table}

\subsubsection{Applicability To Dynamic Systems}\label{subsec:eval_delta}

All the noises discussed in our system have been applied to the plain simulated timeseries $x^*(t)$ as an additional value $n_i(t)$ to generate a signal $\hat{x}^*(t)$ that approximate the real process $\hat{x}(t)$, as discussed in Section~\ref{sec:case}.  
However, the mechanisms we employed to add the noise are applicable in real-time and therefore could be used in dynamic simulators that support user interaction and consequent change in the simulated process. 
\looseness = -1

For instance, a developer may allow the attacker to modify the value of a coil in a power grid to open or close a switch. This will disconnect or connect, respectively, a load that has an impact on the simulated process---it will require additional current or release the used current, respectively. This dynamic behavior is a key concept of a high-interaction honeypot, and the \newframework{} works adequately even in these conditions as shown by the results of the experiment described next.

We modified the developed simulations discussed in Section~\ref{subsec:simulations} to include the effect of opening or closing a switch. We connected a load requesting $4A$ or disconnected a load absorbing $3A$. This does not directly impact the voltage but has a proportional effect on the powers. Then, we employed this new data for testing on our previously trained model (i.e., without including these new signals in the training set) and measured how effective the noises were in these new conditions. 

Results are shown in Figure~\ref{fig:sim_scores} for V1 and I1 (orange bars). I1 is the most influenced measure in this scenario since it is the variable directly influenced by the load change. We see how the Gaussian noise performs better, indicating that our framework can successfully classify the noise instead of the underlying physical process. We also note how probabilities are higher with the fluctuation scenario, but the order is kept almost unchanged.
In the case of V1, \ac{gmm} remains a good representation of the noise with similar probabilities. However, the high increase in the autoencoder suggests that such a generative approach is best suited for dynamic solutions that require more adaptability.

We summarized the probability difference between each noise applied to the normal system and the dynamic scenario in Table~\ref{tab:best_noises_summary} in the \emph{Delta} column. We can see the low probability changes, especially for the voltage and current cases. These results show how \newframework{} is robust to dynamic changes in the system, making it applicable in high-interaction honeypots.

\section{Related Work}\label{sec:related}

After the infamous Stuxnet attack in 2010~\cite{falliere2011w32}, awareness of \ac{ics} security started to rise, and academia and industry began developing and deploying decoy systems to trick the attacker into revealing their attack techniques without compromising the real target. These systems, called ICS honeypots, exhibit different capabilities, from low interaction honeypots~\cite{provos2003honeyd} to more advanced architectures~\cite{salazar2024icsnet,lucchese2023honeyics,conti2022icspot,lopez2020honeyplc}.
While low interaction honeypots are easy to identify~\cite{zamiri2019gas}, researchers also propose solutions to detect high interaction ones~\cite{srinivasa2023gotta}.

An interesting approach has been taken by HoneyJudge~\cite{zhu2024honeyjudge}.
This framework considers six different parameters to identify real or simulated \acp{plc}. They relate to memory handling, logic execution, and the behavior of the physical process. In particular, the authors propose collecting data from the I/O memory of \acp{plc} to extract features such as sensor noise distributions and process dynamics. 
HoneyJudge employs physical equations to identify the right time for the data collection from the \ac{plc} under test. which needs to be accessible during the profiling. Our solution, instead, does not require equations of the physical process and works on systems that could be slightly different from the real one.

In~\cite{ahmed2017hardware}, the authors propose a defensive mechanism against \ac{ics} physical attacks. The system can detect swapping or replacement of sensors by extracting a fingerprint from each sensor noise. It extracts the noise by modeling a simple water tank as an LTI system and using the model to extract the noise from the measurements. 
A similar solution has been proposed by Ahmed et al.~\cite{ahmed2018noise}. They presented a way to fingerprint sensors based on imperfections and noise. The authors proposed to use this information to detect sensor spoofing attacks. The system works by employing subspace system identification techniques to identify the physical equations governing the system from a dataset of measurements from all the sensors. However, that methodology is suited for LTI systems such as the water treatment system employed in the paper. Still, it does not work properly on dynamic (non-LTI) systems such as a power grid. Other defensive mechanisms have been proposed, alleviating the requirements for differential equations of the system~\cite{luo2021deepnoise, aoudi2018truth}. However, no one ever used noise as a source to benchmark simulation fidelity in the \ac{ics} context.

Fuzzy testing has been proposed to identify honeypots~\cite{sun2020identifying}. It collects device behaviors through a series of packets used as probes and feeds their responses to a deep learning classifier to detect simulated environments. %
While this system seems compelling, it represents an orthogonal research line to ours. The authors' solution fingerprints a lack of emulation of industrial devices and their protocols but does not consider the underlying physical process. Moreover, solutions exist to prevent the fuzzy identification of honeypots~\cite{xu2024novel}. 

Different works discussed the characterization of the noise present in a physical process, particularly in power systems~\cite{dehra2018characterization}. Many papers characterize the measurement noise as a Gaussian noise~\cite{brown2016characterizing}. However, other research investigates the presence of other non-Gaussian noises, such as the Laplacian and Gaussian Mixture noise discussed in~\cite{vcubonovic2024impact, martinez2020analysis}.

\section{Conclusion}\label{sec:conclusions}

In this paper, we introduced \newframework{}, a novel framework to assess the physical process fidelity of an \ac{ics} simulation based on the underlying noise of the system and its components. The framework is general purpose and can be adapted to different scenarios, without any assumption on the system to simulate (e.g., no mathematical equations describing the system are needed). We propose a case study based on the EPIC testbed and derive different simulations of its physical process, characterized by the presence of different kinds of noise. After a fine-tuning of the framework on our dataset, we employ \newframework{} and find out that the simulations employing noise from Gaussian and \ac{gmm} distributions were the most similar to the real scenario, together with the noise generated by the autoencoder. Moreover, we show how different measures could be characterized by different noises, and our system allows benchmarking different noises to select the most realistic one to be employed in the simulated scenario under development. Finally, we show how \newframework{} is resistant to process variations and keeps its capabilities of identifying the most suited noise even in the presence of dynamic systems. 

\paragraph{Limitations.}
Despite a quite good amount of \ac{ics} datasets available in the literature~\cite{conti2021survey}, only a few consist of power grid simulations providing physical measurements~\cite{adepu2019epic, pan2015developing}, while others only contain network traffic~\cite{gomez2019generation}. Moreover, not all the testbeds have sufficient documentation allowing for the generation of realistic replicas~\cite{pan2015developing}. 
In other scenarios, such as water treatment, the situation is not much better. In general, while simulators are more widely used, data from real systems is still scarce. Additionally, our simulations do not perfectly reflect the entire EPIC testbed in all its components. This limitation arises from the complexity of power grids and cyber-physical systems in general, as well as the limited number of tools available to efficiently simulate a real-world scenario. Furthermore, details regarding testbeds and datasets~\cite{conti2021survey} are often lacking, making it difficult to accurately emulate a real system.
\looseness = -1

\paragraph{Future Work.}
Testing our \newframework{} on new datasets shared by the community in the coming years is an interesting path for future research. We also invite researchers who are building testbeds and datasets to share simulations and detailed descriptions of their systems, as this will enable the community to perform research similar to this one. Another promising direction for future work involves exploring the problem from an attacker’s point of view. Developing noise fingerprinting systems could allow for the identification of simulations. This would pave the way for the development of anti-honeypot fingerprinting techniques, which may enhance the current state of the art in the field~\cite{srinivasa2023gotta}. Finally, new and more complex generative solutions to add noise to simulations can be investigated, even employing physics-aware generative
models~\cite{zubatiuk2021development}.

\begin{acks}
Work partially funded by the European Union under grant agreement no. 101070008
(ORSHIN project). Views and opinions expressed are however those of the
author(s) only and do not necessarily reflect those of the European
Union. Neither the European Union nor the granting authority can be held
responsible for them. Daniele Antonioli  has been partially supported by the French
National Research Agency under the France
2030 label (NF-HiSec ANR-22-PEFT-0009).
Massimo Merro has been partially supported by the SERICS project (PE00000014) under the \textit{MUR National Recovery and Resilience Plan}, funded by the EU - NextGenerationEU.
\end{acks}

\bibliographystyle{ACM-Reference-Format}
\bibliography{bibliography}

\appendix

\section{Appendix}

\subsection{tsfresh Relevant Features}\label{app:features}
In this section, we list the features from \texttt{tsfresh}~\cite{tsfresh} we kept while extracting features from the original signal that extract information from the noise variation and not from the underlying process.
\begin{itemize}
    \item $approximate\_entropy(x, m, r)$. Implements a vectorized Approximate entropy algorithm.
    \item $kurtosis(x)$. Returns kurtosis of $x$ calculated with the adjusted Fisher-Pearson standardized moment coefficient $G2$.
    \item $lempel\_ziv\_complexity(x, bins)$. Calculate a complexity estimate based on the Lempel-Ziv compression algorithm.
    \item $longest\_strike\_above\_mean(x)$. Returns the length of the longest consecutive subsequence in x that is bigger than the mean of x.
    \item $longest\_strike\_below\_mean(x)$. Returns the length of the longest consecutive subsequence in x that is smaller than the mean of x.
    \item $number\_peaks(x, n)$. Calculates the number of peaks of at least support n in the time series x.
    \item $permutation\_entropy(x, tau, dimension)$. Calculate the permutation entropy.
    \item $skewness(x)$. Returns the sample skewness of $x$ calculated with the adjusted Fisher-Pearson standardized moment coefficient $G1$.
    \item $autocorrelation(x, lag)$. Calculates the autocorrelation of the specified lag.
\end{itemize}

\end{document}